\documentclass[twocolumn,showpacs,preprintnumbers,amssymb,pre,superscriptaddress]{revtex4}

\usepackage{graphicx}
\usepackage{dcolumn}
\usepackage{bm}

\begin{document}


\title{Fluid-membrane tethers: minimal surfaces and elastic boundary layers}

\author{Thomas R. Powers}
 \email{Thomas_Powers@brown.edu} \affiliation{Division of
Engineering, Brown University, Providence, RI 02912}
\author{Greg Huber}%
 \email{huber@umb.edu}
\affiliation{%
Department of Physics, UMass Boston, Boston, MA 02125
}%

\author{Raymond E. Goldstein}
\email{gold@physics.arizona.edu}
\affiliation{Department of Physics and Program in Applied
Mathematics, University of Arizona, Tucson, AZ  85721
}%

\date{September 16, 2001; revised, December 5, 2001}

\begin{abstract}
Thin cylindrical tethers are common lipid
bilayer membrane structures, arising in situations 
ranging from
micromanipulation experiments on artificial vesicles to the
dynamic structure of the Golgi apparatus.  We study the 
shape and formation of a tether in terms of the classical 
soap-film problem, which is applied to the case of a 
membrane disk under tension subject to a point force. 
A tether forms from the elastic boundary layer near the point
of application of the force, for sufficiently large displacement.
Analytic results for various aspects of the membrane 
shape are given.

\end{abstract}


\maketitle

\section{Introduction}

Imagine a soap film connecting two nearby parallel rings which are
slowly pulled apart. The film evolves through a series of catenoids 
until the ring separation reaches a critical value. Then
the film breaks.  Now imagine performing the same experiment with
microscopic rings connected by a lipid bilayer membrane.  In this
paper we will show that the membrane forms catenoidal shapes at
small ring separations, but instead of breaking, the membrane
forms a thin cylindrical tether for sufficiently large
displacement.

A situation very much like this thought experiment arises in a
host of real experiments on artificial vesicles, living cells,
and organelle membranes such as that of the Golgi apparatus. 
Perhaps the
most controlled tether experiment is that of Evans and Yeung, in
which a tether forms when a micropipet is withdrawn from a
spherical vesicle held at a fixed tension and bonded to a
stationary bead~\cite{EY}. There are many variations on this
experimental theme~\cite{HMB,Waugh,HWEM,HSDS,HBSZ}, and such 
experiments
have been used to measure a wide variety of membrane mechanical
properties~\cite{dai_sheetz}.

\begin{figure*}
\includegraphics[height=2in]{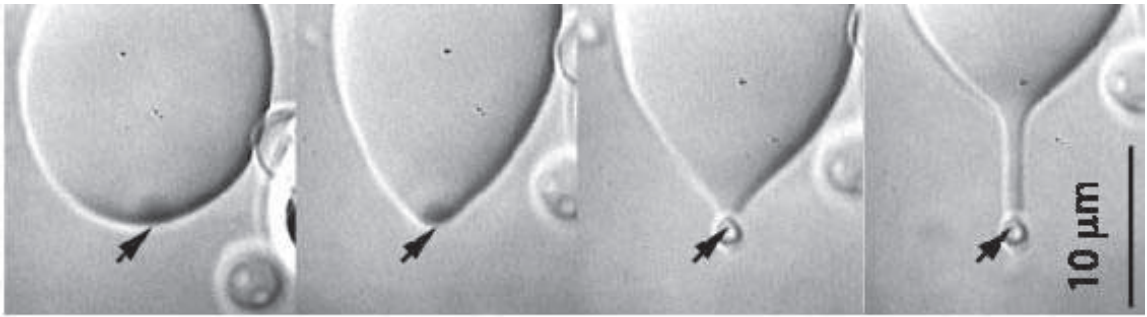}
\caption{Equilibrium shapes of a vesicle subject to various point
forces.  An optical tweezer exerts the point force, which
increases from left to right. Figure courtesy of D. Fygenson.}
\label{deborahpicture}
\end{figure*}

Tethers commonly form in less controlled situations as well.
Simply pulling on a vesicle or cell with a sufficiently large
point force leads to a membrane tether
(Fig.~\ref{deborahpicture}). Tethers form when tubulin trapped
inside a vesicle polymerizes to form microtubules~\cite{MH}: as
the microtubules grow, the initially spherical membrane at first
distorts into an ellipsoidal shape, and then eventually forms a
surface of revolution with  a contour in the shape of the Greek
letter ``$\phi$"~\cite{EFL,FML,ECT}.  Improvements in staining
techniques have recently revealed dynamic tether networks in the
Golgi apparatus  of living cells~\cite{JLS}; similar model
membrane networks have been studied \textit{in vitro}
and used as templates for making more durable
networks~\cite{EvansSci}.

\begin{figure}
\includegraphics[height=2.718in]{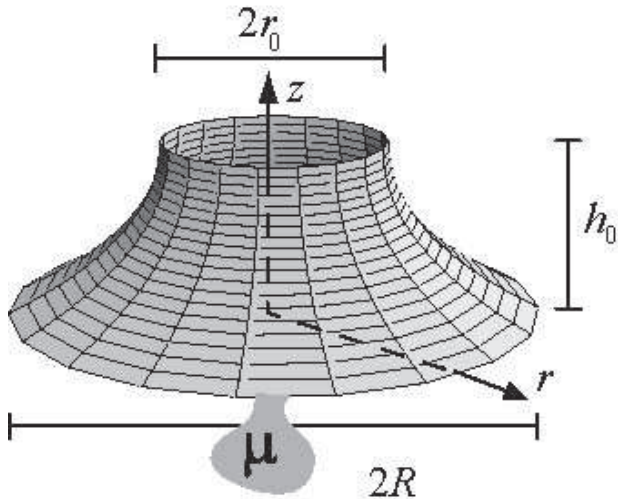}
\caption{Model problem.  Two parallel rings with aligned centers
are connected by a lipid bilayer membrane in contact with a lipid
reservoir at fixed chemical potential (per unit area) 
 $\mu$.} \label{model}
\end{figure}

The variety of conditions under which tethers occur shows that they
are robust structures, insensitive to the details of applied
forces and boundary conditions. Despite much theoretical, 
computational, 
and experimental attention, a simple intuitive picture for 
tether formation
which exploits the insensitivity to geometric constraints 
(such as volume conservation) and emphasizes the role of 
the most essential
control parameters has yet to emerge. 
It is therefore natural to study the theory of
tethers in the simplest possible situation, namely
the classic soap film geometry described above (Fig.~\ref{model}).
This model situation captures the essential features of the tether
shape and formation without the experimentally important, but
ultimately complicating, effects of volume and lipid
conservation.  Our quasi-analytic approach exploits the smallness
of the tether radius, and is complementary to important numerical
work by various groups, most notably  Heinrich \textit{et
al.}~\cite{HBSZ}.
We begin our analysis in section II with a discussion of shells,
balloons, and soap films in order to contrast their familiar
mechanical properties with the peculiar properties of lipid
bilayer membranes.  The latter are the subject of section III,
which defines precisely the model problem.  Section IV reviews
the elements of the soap film problem which are relevant for
understanding the formation of tethers, the subject of
section V.  There we use asymptotic methods to solve the
linearized equations for small ring displacements and large
tensions, and find that the membrane shape is that of a catenoid
with a small elastic boundary layer surrounding the smaller ring
(the point force).  At a critical ring separation, a tether forms
from the elastic boundary layer.   The tether shape is studied
analytically and numerically, and a connection is made with 
the classic film-coating calculations of Landau and
Levich~\cite{LL}, as well as Bretherton's related calculation of
the shape of a long air bubble rising in a fluid-filled capillary
tube~\cite{B}.  Section VI. is the conclusion. The appendix reviews
the subtleties that arise when comparing the variational approach
for lipid bilayer membrane elasticity to that of moment and force
balance.

\section{Shells, Balloons, and Soap Films}
\label{shells}

Before exploring the lipid bilayer membrane properties that give
rise to tethers, we review three examples of elastic surfaces
encountered at the macroscopic scale: shells, balloons, and soap
films.  Much intuition about elastic surfaces derives from
these canonical examples. Some of this intuition can be directly
applied to lipid membranes, but it is illuminating to point out
the crucial differences as well.

Shells are solid surfaces with a small thickness and a preferred
shape in the absence of external stresses. A plate is a shell with
a flat preferred shape. It is a matter of common experience that
the force required to bend a thin plate through a certain
displacement is much less than that required to stretch the plate
through the same displacement.  The plate can undergo large
displacements through bending without subjecting each element to
large stress; thus, linear elasticity is valid and the
nonlinearities in the equations for shape are solely geometrical.
In the limit of small plate thickness, stretching energy exceeds
bending energy, since bending is differential stretching.
Deformations without stretching are therefore of lowest 
energy~\cite{LLelasticity}. These deformations, in which 
the distances between nearby points remain fixed,
are called {\it isometric}. 
For
example, the axisymmetric isometric deformations of the plane are
cylinders and cones~\cite{dgbook}.  Note that a shearing motion
locally stretches the plate, even if the total area does not
change. Bending energy (and possibly boundary conditions) removes
the degeneracy when, as in the case of a flat plate, there is a
multiplicity of isometric deformations. Bending also plays a role
near boundaries and point forces.  A familiar example of a plate
with vanishing thickness is a sheet of paper, which bends easily
but hardly stretches or shears. The reader can easily verify that
a sheet of paper subject to a point force forms a cone with a
single fold, a shape with two-fold symmetry.  
For shells, the isometric constraint  on
deformations is even more severe.  \textit{Any} deformation of a
sphere requires some local stretching or shearing:  no
almost-spherical shapes are isometric to the sphere. A spherical
shell is \textit{geometrically rigid}.  On the other hand, there
are isometric small deformations of a spherical shell with a
circular hole~\cite{LLelasticity}.  The problem of determining the
geometric rigidity of an axisymmetric shell with boundaries was
solved using the qualitative theory of differential
equations~\cite{geomrig}; for recent progress using a geometric
approach, which can be generalized to non-axisymmetric shells,
see~\cite{audoly}.

Next, consider the case of balloons.  Like plates with vanishing
thickness, there is virtually no cost to bending a section of a
balloon compared to that of stretching or shearing.  However,
deformations are not required to be isometric since the cost of
stretching is also low.  Such a material is called a ``membrane"
in the theory of elasticity~\cite{LLelasticity}; to avoid
confusion with lipid bilayer membranes, this
terminology will not be used.  
Since real balloons are easily stretched out of the
linear elastic regime, the nonlinearity of the governing equations
arises from the constitutive relations as well as the geometry of
large deformations. Anyone who has inflated a cylindrical balloon
has seen a phenomenon very much like the tethers of
Fig.~\ref{deborahpicture}.  A partially inflated balloon has two
cylindrical regions, one smaller than the other, which are
smoothly connected by a junction region.
However, this phenomenon differs fundamentally from that of tether
formation in lipid bilayer membranes for several reasons.  No
point force is required to make a balloon form a tether;  it arises
from the nonlinear constitutive relation between tension and areal
extension, which is reflected by a non-monotonic $p$-$V$ curve,
reminiscent of the isotherms of the van der Waals equation of
state for a gas~\cite{hutch}. The material in the larger cylinder
is stretched more, and thus has a higher tension. The smaller
cylinder is under smaller tension; the difference between the
axial force of each tension is precisely balanced by the net
pressure on the junction region. Thus, a pressure jump across the
surface of the balloon is crucial for this tether. Below it is 
shown that such a pressure jump is unnecessary for lipid bilayer
membrane tethers.

Our final canonical example is the ``ideal" soap film, which simply 
minimizes area subject to the
boundary conditions and volume constraints. Effects
such as thickness variations and draining
do not have a clear counterpart in the case of lipid bilayer
membranes, and are therefore disregarded.  
Since soap films are liquid, the static in-plane shear
rigidity vanishes, and the notion of geometric rigidity does not
apply. Deformations need not be isometric and folds are not
required to bend a flat film into a hemisphere.  (There can be
solid-like folding behavior in liquid film dynamics whenever
bending flows are preferable to extensional flows~\cite{maha}.)
Unlike the case of plates, interfacial tension is a material
property of the soap film, and is not determined by external
forces. To imagine poking a soap film with a point force, consider
again the geometry of two rings described in the introduction, but
now
take one ring radius to be much smaller than the other. 
Equilibrium solutions exist only when the ring separation is less
than a critical separation, comparable to the radius of the
smaller ring. At a slightly larger separation, the film breaks.
The equilibrium shapes are shallow catenoids, and look nothing
like tethers (structures reminiscent of thin tethers form during
the rupture of a soap film, but these have a dynamical origin;
see~\cite{steen}). However, these shallow catenoidal shapes resemble
the junction region between the tether and vesicle of
Fig.~\ref{deborahpicture}.  Soap films and catenoids will be
studied more fully in section IV.

Lipid bilayer membranes have elements of
each of these examples.  Like macroscopic plates, lipid membranes
resist bending, but they are typically fluid and therefore lack
geometric rigidity, like soap films. The interplay between these
solid-like and fluid properties leads to the formation of tethers.

\section{Lipid Bilayer Membranes}

\subsection{Bending elasticity}
There is a vast literature on the mechanics of lipid bilayer
membranes (\textit{e.g.} see~\cite{udo}).   Here we review
only the aspects relevant to tethers. Lipid molecules are 
amphiphilic, composed of two oily hydrophobic chains attached 
to a polar hydrophilic head. Lipids self-assemble in aqueous 
solution to shield the
chains from the water, forming micron-size bilayer surfaces.  In
the fluid phase, a typical self-diffusion constant for a lipid
molecule in a membrane is of order
$10^{-8}$cm$^2$/sec~\cite{alberts}.  The membrane is clearly a
two-dimensional fluid over the time scale of typical tether
experiments:  a molecule will diffuse from one side of a one-micron 
radius spherical vesicle to the other in about a second.
Hence, there is no in-plane static shear modulus.  However,
membranes  have a non-zero bending modulus since bending the
membrane compresses and extends the slightly elastic heads and
tails of the lipid molecules.  The fluid nature of the membrane
constrains the form of the elastic energy to depend only on the
\textit{shape} of the membrane.  To lowest order in curvatures,
the bending energy is given by the expression of Canham and
Helfrich~\cite{canham,helfrich},
\begin{equation}
{\cal E}_{\text{el}}=\int\text{
d}S\biggl[{\kappa\over2}(2H)^2+{\bar\kappa\over2}K\biggr],
\label{can-helf}
\end{equation}
where $H$ is the mean curvature of the membrane surface, and $K$
is the Gaussian curvature ($\kappa$ and $\bar\kappa$ are 
elastic constants).  
In terms of the principal radii of
curvature $R_1$ and $R_2$ at a point on the surface,
$2H=1/R_1+1/R_2$ and $K=1/(R_1R_2)$.  Explicit formulas for these
curvatures will be given below.  For simplicity, suppose that
there is no difference between the two sides of the bilayer;
hence, there is no spontaneous curvature.  The Gaussian curvature
term is typically dropped in studies of vesicles since it amounts
to the sum of a deformation-independent term and a boundary term
by the Gauss-Bonnet theorem:
\begin{equation}
\int\mathrm{d}S K=4\mathrm{\pi}(2-2g)+\oint\mathrm{d}s\ \kappa_g,
\label{gaussbonnet}
\end{equation}
where the genus $g$ is the number of handles and $\kappa_g$ is the
geodesic curvature of the boundary of the surface~\cite{dgbook}.
  Note that a more detailed treatment of the
formation of tethers in vesicles would require additional terms of
the generalized bilayer couple
model~\cite{helfrich1974,evans1974,evans1980,bozic1992,miao1994}.
In the spirit of explaining tether formation in the simplest
possible context, these are disregarded.

Although typical values for the elastic bending modulus $\kappa$
are $10$--$15k_BT$, thermal fluctuations easily excite 
long-wavelength bending modes since the elastic energy 
vanishes as the fourth power of the wavenumber for fluctuations
about a flat sheet. These fluctuations lead to an entropic area
elasticity similar to that of semiflexible polymers, as is most
directly illustrated by the experiments of Evans and
Rawicz~\cite{evans_rawicz1990} (see also~\cite{rawicz_etal2000}).
In these experiments, the tension in a vesicle is measured as a
function of apparent area by suctioning a small amount of the
vesicle membrane into a pipet. At low suctions (or low tensions),
the resistance to stretching is the entropic penalty of reducing
the number of fluctuating modes.  At high tensions, most of the
thermal ripples have been 
smoothed out, and the 
resistance to stretching is mainly due to the membrane's intrinsic
area elasticity. Thermal fluctuations are therefore negligible for the 
high tension regime considered here.

\subsection{Model problem and nondimensionalization}

As in the introduction, consider a lipid membrane spanning two
rings that are initially concentric and lying in the plane $z=0$
(see Fig.~\ref{model}).   The outer ring of radius $R$ remains in
this plane, but the height $h_0$ of the inner ring will be varied.
Measuring all lengths in units of the radius $R$, henceforth $R=1$. 
The inner ring then has radius $r_0\ll 1$. 
Later, we will take $r_0\rightarrow0$ to model the application 
of a point force. 
For comparison with the
case of soap films and also for the numerical approach, it 
is convenient to keep $r_0$ nonzero for now.  Assume that there
is a reservoir of lipid at a fixed chemical potential 
(per unit area) $\mu$. 
Further suppose this ``surface tension" is very large compared to the
bending elasticity, $\mu R^2\gg\kappa$.  Thermal fluctuations are 
therefore negligible. Since the membrane has edges, the Gaussian
modulus $\bar\kappa$ affects the shape through the boundary
conditions (recall the geodesic curvature term of
Eq.~(\ref{gaussbonnet})). For simplicity, we disregard this effect
and set $\bar\kappa=0$.

Our model is closest in spirit to the tether experiments of Evans
and Yeung~\cite{EY} mentioned earlier. The pipet suction sets the
value of the tension $\mu$, and the small amount of lipid
projecting inside the pipet serves as a reservoir. In the
experiments, the vesicle is under pressure and is therefore curved,
whereas in our model there is no pressure jump across the
initially flat membrane. The role of the pressure in vesicle
tethers is small since the tether curvature is much larger than
the vesicle curvature for high tension.  In fact, it is shown 
in section VB2 that the pressure leads to a subleading correction 
to the
tether radius.

The complete specification of the model problem requires defining
the boundary conditions at the rings.  The tangent plane of the
surface at the point force is perpendicular to the force if the
line of action is along the axis of symmetry. Since the role of
the small ring is to mimic a point force, the membrane is clamped
so that the tangent plane at each point on the boundary (with the
small ring) is in the plane of the small ring. The ring at $r=1$ is
somewhat artificial; therefore we choose the simplest possible
boundary condition for $r=1$, which turns out to be zero moment,
$H=0$~\cite{LLelasticity}. The outer ring acts as a hinge.
Defining the dimensionless parameter
\begin{equation}
\epsilon\equiv{\kappa\over \mu R^2}~, \label{epsilondefine}
\end{equation}
(recall $R=1$),
the problem is to minimize the energy (measured in units of $\mu R^2$) 
\begin{equation}
{\cal E}=\int\mathrm{d}S+{\epsilon\over2}\int\mathrm{d}S(2H)^2
\label{variationaleq}
\end{equation}
for given ring separation $h_0$ subject to the boundary conditions
and $\epsilon\ll1$. Equation~(\ref{variationaleq}) casts the
tether problem into the same form as the classic variational problem 
for a minimal surface ($\epsilon=0$).

\subsection{Euler-Lagrange equations}

The derivation of the Euler-Lagrange equations from the energy of
Eq.~(\ref{variationaleq}) is somewhat lengthy but
straightforward~\cite{OYHel}.  The result is
\begin{equation}
2{\epsilon}({\nabla^2}H+2H^3-2HK)-2H+\Delta p=0, \label{eleqn}
\end{equation}
where $\nabla^2$ is the covariant Laplacian on the surface.
$\Delta p$, measured in units of $\mu/R$ (in other words, $\mu$) 
is zero for our soap
film geometry, but it is included in Eq.~(\ref{eleqn}) for later
discussion of the effect of pressure on tether shape. Note that
$\epsilon=0$ and $\Delta p=0$ yields the minimal-surface equation,
$H=0$. Since the membrane shape is a surface of revolution,
natural coordinates for the surface are $\varphi$, the azimuthal
angle, and $s$, arclength along a meridian.  Arclength is
measured from the inner ring, which has coordinate $s=0$.  The
position of a point on the surface is therefore $\mathbf
{X}(s,\varphi)=r(s)\mathbf{\hat r}+z(s)\mathbf{\hat z}$, where $r$
and $z$ are cylindrical coordinates.  Note that $r_s^2+z_s^2=1$,
since $s$ is arclength.

With these choices, the metric, or first fundamental form, is
\begin{equation}
g_{ij}\mathrm{d}\xi^i\mathrm{d}\xi^j=\mathrm{d}s^2
+r^2\mathrm{d}\varphi^2,
\label{firstff}
\end{equation}
where $\xi^1=s$ and $\xi^2=\varphi$. The second fundamental form
is
\begin{equation}
K_{ij}\mathrm{d}\xi^i\mathrm{d}\xi^j=(z_sr_{ss}-r_sz_{ss})\text{
d}s^2-rz_s\mathrm{d}\varphi^2, \label{secondff}
\end{equation}
where  $z_s=\mathrm{d}z/\mathrm{d} s$, \textit{etc.}  We follow
the usual conventions for raising and lowering indices using the
inverse $g^{ij}$ of the metric tensor.  Thus,
$g^{ik}g_{kj}=\delta^i_j$, $K^i_j=g^{ik}K_{kj}$, and
\begin{eqnarray}
H&\equiv&{1\over2}g^{ij}K_{ij}={1\over2}\bigg[{r_{ss}\over
z_s}-{z_s\over r}\bigg],\label{meancurv}\\
K&\equiv&\det K^i_j=-{r_{ss}\over r},\label{gausscurv}\\
\nabla^2&\equiv&{1\over\sqrt
g}\partial_ig^{ij}\sqrt{g}\partial_j={1\over
r}{\mathrm{d}\over\mathrm{d}s}r{\mathrm{d}\over\mathrm{d}s},\label{laplacian}
\end{eqnarray}
where $g$ is the determinant of the metric tensor $g_{ij}$, and
$r_s^2+z_s^2=1$ was used to simplify~(\ref{meancurv}).

The Euler-Lagrange equation~(\ref{eleqn}) for an axisymmetric
shape is a nonlinear ordinary differential equation, easily solved
with numerical methods.  However, more insight is gained by
exploiting the smallness of $\epsilon$. Since $\epsilon$ multiplies the
term in~(\ref{eleqn}) with the highest number of derivatives, there will be
a boundary layer.   The layer occurs where the boundary conditions are
incompatible with Eq.~(\ref{eleqn}) with $\epsilon=0$.  It is now
clear why the zero moment $H=0$ boundary condition is the
most natural condition at the larger ring: this boundary condition
is compatible with the minimal-surface equation and does not lead
to a boundary layer at the larger ring.  The clamped boundary
condition at the smaller ring, however, is incompatible with the
minimal-surface equation, since the ring must exert a moment on
the surface to keep it clamped. In this boundary layer, bending and
tension balance and the shape is determined by the full
Euler-Lagrange equation, where the smallness of $\epsilon$ is
offset either by large $s$-derivatives or large curvatures. 
Balancing $\epsilon\nabla^2 H$ with $H$ in~(\ref{eleqn}) 
reveals that the boundary layer thickness scales as $\sqrt{\epsilon}$.
In the outer region, beyond the
elastic boundary layer, bending is unimportant and the shape is
governed by the minimal-surface equation, $H=0$.  The only
nonplanar axisymmetric minimal surface is a catenoid, the surface
of revolution generated by a catenary~\cite{dgbook}.  Thus,
the membrane forms a catenoid in the outer region.  In the next
section, reviews basic facts about catenoids.

\section{Catenoid Lore}

We noted in the previous section that setting $\epsilon=0$ and
$\Delta p=0$ in Eq.~(\ref{eleqn}) results in the minimal-surface
equation $H=0$. To leading order, the outer solution to
Eq.~(\ref{eleqn}) is given by exactly the same condition. The
solution to the minimal-surface equation 
is conveniently found by applying Noether's theorem directly to
the energy functional~(\ref{variationaleq}) with $\epsilon=0$. To
this end, rewrite $\mathrm{d}S$ in terms of $r(z)$:
\begin{equation}
{\cal E}=\int\mathrm{d}\varphi\mathrm{d}z \ r\sqrt{1+r_z^2}.
\label{area_energy}
\end{equation}
The conserved quantity associated with the invariance of the
integrand of~(\ref{area_energy}) with respect to translations in
$z$ is the axial force $F$ necessary to hold the rings apart at 
a given separation:
\begin{equation}
{F\over2\mathrm{\pi}}={r\over\sqrt{1+r_z^2}}. \label{Noether}
\end{equation}
The radius $r$ attains its minimum value $b=F/(2\mathrm{\pi})$
when $r_z=0$. Integrating~(\ref{Noether}) yields the catenoid
\begin{equation}
r=b\cosh\bigg({z-c\over b}\bigg), \label{catshap}
\end{equation}
where $c$ is the $z$-coordinate of the minimum radius. Note that
$b$ is the minimum \textit{possible} radius; \textit{i.e.} it is
possibly \textit{not} attained. The minimum radius is attained
only if $c\le h_0$.
\begin{figure}
\includegraphics[height=2.in]{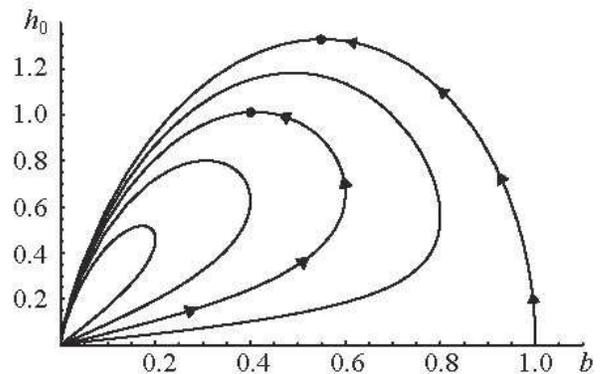}\caption{Ring
separation $h_0$ \textit{vs.} minimum neck radius $b$ (force by
$2\mathrm{\pi}$) for the values $r_0=0.2, 0.4, 0.6, 0.8$, and
$1.0$, proceeding from the innermost curve to the outermost curve.
For a given $r_0$, as the rings are separated, the sequence of
shapes corresponds to the trajectory marked with arrows.}
\label{cat_h0_v_b_fig}
\end{figure}

\begin{figure}
\includegraphics[height=3.in]{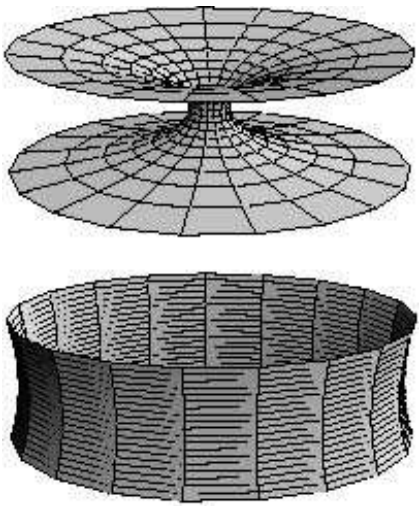}\caption{The
two catenoidal solutions with equal-size rings ($r_0=1$) and ring
spacing $h_0=0.6$; the lower catenoid has less area.}
\label{catenoid_fig}
\end{figure}

The boundary condition $r(0)=1$ determines $c=\pm\cosh^{-1}(1/b)$,
and thus Eq.~(\ref{catshap}) expands to
\begin{equation}
r=\cosh\bigg({z\over b}\bigg)\mp\sqrt{1-b^2}\sinh\bigg({z\over
b}\bigg). \label{catshapeII}
\end{equation}
The upper sign is chosen as it corresponds to catenaries with a
minimum neck radius at a positive value of $z$ ($c>0$). The force
$F$, or equivalently the minimum neck radius $b$, is determined by
the boundary condition at the other ring: $r(h_0)=r_0$, or
\begin{equation}
r_0=\cosh\bigg({h_0\over b}\bigg)-\sqrt{1-b^2}\sinh\bigg({h_0\over
b}\bigg). \label{fixc}
\end{equation}
Solving~(\ref{fixc}) for $h_0$ gives the separation as a function
of force:
\begin{equation}
h_0=b\log\Bigg({r_0\pm\sqrt{r_0^2-b^2}\over1-\sqrt{1-b^2}}\Bigg).
\label{cat_sepvforce}
\end{equation}
Note that the two branches form a closed curve in the $b$-$h_0$
plane for $r_0<1$ (Fig.~\ref{cat_h0_v_b_fig}). Since each curve
has a maximum (marked with a dot for the curves $r_0=0.6$ and
$r_0=1.0$), there is a critical $r_0$-dependent separation beyond
which no catenoidal solution exists. The soap film spanning the two
rings breaks just beyond this critical separation.
For fixed $r_0$ and a given separation $h_0$ below the maximum,
there are two catenoidal solutions. For example,
Fig.~\ref{catenoid_fig} illustrates the two equilibrium catenoids
with $r_0=1$ and $h_0=0.6$.
For a given $h_0$, one can show that the catenoid with the larger
$b$ has less area.  Thus, the solution with the smaller neck
(\textit{e.g.} the upper catenoid in Fig.~\ref{catenoid_fig}) is
unobservable in real soap films.  In the presence of bending
stiffness and a fully-developed tether, it is shown below that
the axial force is $2\mathrm{\pi}\sqrt{2\epsilon}$.  To see which
catenoid matches onto a fully developed tether, consider the
extreme case $r_0=1$.  Since the axial force vanishes as
$\epsilon\rightarrow0$, the matching catenoid must have $b$ near
zero and lie on the left branch of the $r_0=1$ curve of
Fig.~\ref{cat_h0_v_b_fig}. By continuity, as $r_0$ decreases, the
corresponding matching catenoid is always on the left branch.
Therefore, the catenoid joining a tether is the one with a narrow
neck; bending stiffness selects this otherwise unobservable shape.

Finally, the arrows in Fig.~\ref{cat_h0_v_b_fig} display the
trajectory of shapes as the rings are separated.  The force $F$
increases from zero to a maximum value, and then decreases
slightly before the film breaks. This
nonmonotonic behavior occurs in the case of tethers as well, as 
is shown below.

\section{From Catenoids to Tethers}

As discussed earlier, for small but nonzero $\epsilon$, the outer
region of the membrane will have a catenoidal shape, and there will
be an elastic boundary layer near the small ring.  This elastic
boundary layer allows the attainment of the point force limit,
$r_0\rightarrow0$, in contrast to the case of the soap film. We
begin with the analysis of the case of small axial separation,
$h_0\ll1$.

\subsection{Small displacements}

When $h_0\ll1$, it is most convenient to work in the Monge
parameterization, in which the surface is represented by its height
$z(r)$ above the plane $z=0$.  To leading order in $h_0$,
$s\approx r$, the mean curvature $H\approx(1/2) \nabla^2 z$, and the 
Euler-Lagrange equation reduces to
\begin{equation}
\epsilon\nabla^4 z-\nabla^2 z=0, \label{lin_eleqn}
\end{equation}
with
\begin{equation}
\nabla^2={1\over r}{\mathrm{d}\over\mathrm{d}r}\bigg[
r{\mathrm{d}\over\mathrm{d}r}\bigg]. \label{nabla_lin}
\end{equation}
The boundary conditions at the large ring $r=1$ are $z=0$ and the
condition of zero moment, $\nabla^2z=0$. At the inner ring
$r=r_0$, the displacement $z=h_0$ and the slope
$\mathrm{d}h/\mathrm{d}r=0$.

At order $\epsilon^0$, the outer solution satisfies $\nabla^2
z_{\text{outer}}(r)=0$, \textit{i.e.}
\begin{equation}
z_{\text{outer}}(r)=b_1+b_2\log r. \label{outer_soln}
\end{equation}
The boundary condition on displacement at $r=1$ fixes $b_1=0$. The
zero-moment boundary condition adds no constraint on the solution
of~(\ref{outer_soln}); $b_2$ must be determined by matching to the
inner solution.

Note that the outer solution diverges at the inner ring as
$r_0\rightarrow0$; the inner solution must correct for this
divergence. To find the inner solution, expand the region near
$r=0$ with the rescaling $\rho=r/\sqrt{\epsilon}$. Then the inner
solution satisfies
\begin{equation}
{1\over
\rho}{\mathrm{d}\over\mathrm{d}\rho}\bigg(\rho{\mathrm{d}\over\text{
d}\rho}\bigg)\bigg[{1\over
\rho}{\mathrm{d}\over\mathrm{d}\rho}\bigg(
\rho{\mathrm{d}\over\mathrm{d}\rho}\bigg)z_{\text{inner}}+z_{\text
{inner}}\bigg]=0, \label{outer_eleqn}
\end{equation}
or
\begin{equation}
{z_{\text{inner}}(\rho)\over h_0}=c_1+c_2\log
\rho+c_3I_0(\rho)+c_4K_0(\rho), \label{inner_soln}
\end{equation}
where $I_0(\rho)$ and $K_0(\rho)$ are modified Bessel functions.
$c_3$ must vanish, since $I_0(\rho)$ diverges as
$\rho\rightarrow\infty$ and cannot be matched to the outer
solution. The boundary conditions at $r=r_0$ 
(\textit{i.e.}, $\rho=r_0/\sqrt{\epsilon}$) 
add two more constraints to yield
\begin{eqnarray}
{z_{\text{inner}}(\rho)\over h_0}
=1+&c_4&\bigg[K_0(\rho)-K_0\bigg({r_0\over\sqrt{\epsilon}}\bigg)\nonumber\\
&+&{r_0\over\sqrt{\epsilon}}
K_1\bigg({r_0\over\sqrt{\epsilon}}\bigg)\log\bigg({\rho\sqrt{\epsilon}\over
r_0}\bigg)\bigg]. \label{inner_solnII}
\end{eqnarray}

To match the inner and outer solutions, note that $K_0(\rho)$
decays exponentially at large $\rho$.  Therefore, the constant
terms of $z_{\text{inner}}$ must vanish, and the coefficient of
the logarithmic term of $z_{\text{inner}}$ must match $b_2$. To
leading order, 
\begin{equation}
{z_{\text{outer}}\over h_0}={
{r_0\over\sqrt{\epsilon}}K_1({r_0\over\sqrt{\epsilon}})\log r\over
{r_0\over\sqrt{\epsilon}}K_1({r_0\over\sqrt{\epsilon}})\log
r_0+K_0({r_0\over\sqrt{\epsilon}})}, \label{out_final}
\end{equation}
and
\begin{equation}
{z_{\text{inner}}\over h_0}={
{r_0\over\sqrt{\epsilon}}K_1({r_0\over\sqrt{\epsilon}})\log
r+K_0({r\over\sqrt{\epsilon}})\over
{r_0\over\sqrt{\epsilon}}K_1({r_0\over\sqrt{\epsilon}})\log
r_0+K_0({r_0\over\sqrt{\epsilon}})}. \label{inner_final}
\end{equation}
The Bessel function $K_0(r/\sqrt{\epsilon})$ cancels the
logarithmic divergence of the $\log(r)$ term of the inner solution.
To construct a uniformly-valid approximation
$z_{\text{composite}}(r)$ for both the inner and outer regions,
add the two solutions and subtract their common part~\cite{vandyke}. 
This
procedure yields the very compact result
\begin{equation}
z_{\text{composite}}=z_{\text{inner}}~. \label{matching_final}
\end{equation}
\begin{figure}
\includegraphics[height=1.5in]{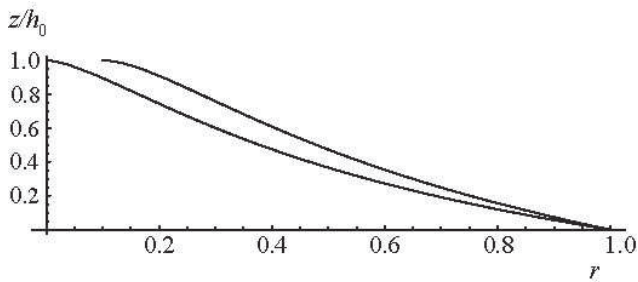}
\caption{Comparison of membrane profiles 
$z_{\text{composite}}(r)$ for $r_0=0.1$ (upper curve) and the point 
force limit $r_0\rightarrow0$
(lower curve).}
\label{compare_fig}
\end{figure}

It is now clear why a small amount of bending elasticity allows a
membrane to support a point force: the elastic boundary cuts off
the divergence of the logarithm of the outer solution.  To find
the solutions in the limit of a point force, recall
$K_1(r_0/\sqrt{\epsilon})=\sqrt{\epsilon}/r_0+{\cal
O}(r_0/\sqrt{\epsilon})$ and $K_0(r_0/\sqrt{\epsilon})=-\gamma
+\log(2\sqrt{\epsilon}/r_0)+{\cal O}((r_0/\sqrt{\epsilon})^2)$,
where $\gamma=0.5772...$ is the Euler constant. Thus
\begin{equation}
z_{\text{composite}}(r)= h_0{\log
r+K_0(r/\sqrt{\epsilon})\over-\gamma+\log(2\sqrt{\epsilon})}.
\label{comp_r0is0}
\end{equation}
Note that the boundary condition $z_{\text{composite}}(1)=0$ is
satisfied up to terms of order
$\epsilon^{1/4}\exp(-1/\sqrt{\epsilon})/\log\epsilon$ for small
$\epsilon$.  Figure~\ref{compare_fig} shows that although the details
of the membrane shape depends somewhat sensitively on the value of $r_0$,
the physical limit $r_0\rightarrow0$ is well behaved.
In this limit, the elastic boundary layer becomes a small disc
of approximate radius $\sqrt{\epsilon}$ around the point force.
Thus, the outer solution is roughly the catenoid that connects a
ring of radius unity with a ring of radius $\sqrt{\epsilon}$. As
in the previous section, the maximum ring separation for a
soap film in this situation is approximately equal to the radius
of the smaller ring.  Thus, as $h_0$ increases, the amplitude of
the catenoid increases until $h_0$ is of order $\sqrt{\epsilon}$.
Since the amplitude of the catenoid cannot increase beyond this
value, the boundary layer deforms into a thin cylinder to
accommodate further increases in $h_0$.  The formation of the
tether is a smooth process; there is no bifurcation.

\subsection{Tether: analytical approach}

\subsubsection{Tether radius, tether stability, and axial force}
Tether formation is an intrinsically nonlinear phenomenon, and to
give a complete account of the tether shape we resort to
numerical methods.  However, many features of the tether yield 
to an analytic approach.  The tether radius is the most prominent 
such feature. Our numerical calculations will verify that
the tether has a cylindrical shape between from the end cap and the
catenoidal junction.  Thus, for our soap-film geometry
with $\Delta p=0$, the radius follows from~(\ref{eleqn}) with
constant mean curvature and vanishing Gaussian curvature:
\begin{equation}
 2\epsilon H^3-H=0.
\end{equation}
Since $H=-1/(2a)$ for a cylinder of radius $a$, the exact tether
radius  $a= \sqrt{\epsilon/2}$~\cite{EY}.  This square-root
dependence of tether radius on inverse tension has been verified
experimentally by Evans and Yeung~\cite{EY}.

To study the stability of a tether, write
$r(z)=a+u(z)$ and expand the elastic energy~(\ref{variationaleq})
to ${\cal O}(u^2)$ (to express the metric~(\ref{secondff}) and
mean curvature~(\ref{meancurv}) as functions of $z$ use
$\mathrm{d}z/\mathrm{d}s=\sqrt{1+r'(z)^2}$), which yields
\begin{eqnarray}
{{\cal
E}\over2\mathrm{\pi}}=\int\mathrm{d}z\bigg[\bigg(a+{\epsilon\over2a}\bigg)
&+&\bigg(1-{\epsilon\over2a^2}\bigg)\bigg(u+{au^{\prime2}\over2}\bigg)\nonumber\\
&+&{\epsilon a\over2}\bigg(u^{\prime\prime2}+{u^2\over
a^4}\bigg)\bigg]. \label{energy_expand}
\end{eqnarray}
A total derivative term has been dropped in~(\ref{energy_expand}).
Minimizing the
$u$-independent terms over $a$ yields the equilibrium tether radius. 
The terms linear in $u$ vanish as
expected when $a$ takes the equilibrium value $\sqrt{\epsilon/2}$.
Note that the terms quadratic in $u^\prime$ vanish in equilibrium
as well, since the terms $u$ and $au^{\prime2}/2$ always enter in
the combination $u+au^{\prime2}/2$. This combination arises from the factor
$r\sqrt{1+r^{\prime2}}$ in the original energy. Since the
remaining terms of Eq.~(\ref{energy_expand}) in $u^2$ and
$u^{\prime\prime2}$ are positive definite, the tether is stable.
Therefore, the \textit{equilibrium} cylinder solution does not
undergo a pearling
instability~\cite{bar-ziv_moses1994,gold_et_al1996}.  These
considerations suggest that the pearling behavior induced by a
rapid pull of a vesicle tether~\cite{bar-ziv_moses_nelson1998}
arises because hydrodynamic resistance prevents the radius from
instantly assuming the value appropriate to the new value of
tension.  This mechanism differs in detail from that of the
laser-tweezer-induced instability of membrane tubes with fixed
volume~\cite{gold_et_al1996}.

\begin{figure}
\includegraphics[height=2.8in]{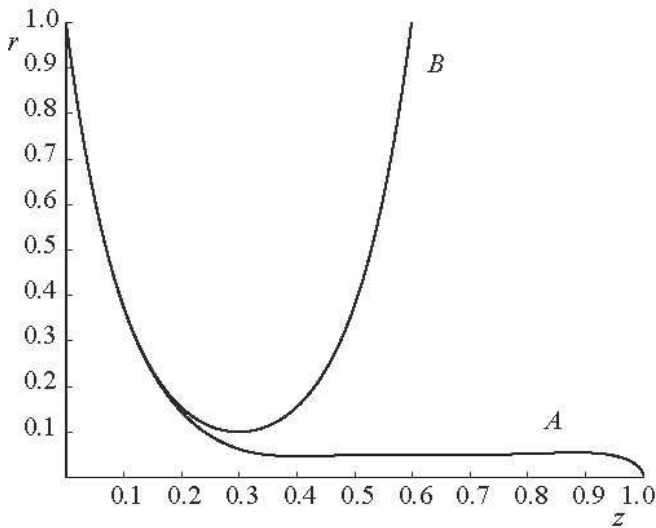}
\caption{Overlay of a fully developed tether ({\it A}) with 
radius $a=\sqrt{\epsilon/2}$ and a catenoid ({\it B}) with minimum radius 
$b=\sqrt{2\epsilon}$. In this example, $\epsilon=0.005$ 
and the transition region lies roughly between  $z=0.2$ and $z=0.4$.  
} \label{overlay_fig}
\end{figure}

The tether radius determines the axial force on the catenoid.   
The argument hinges on the $z$-independence of the axial force.
For an undistorted
cylinder with $u=0$ and equilibrium radius $a=\sqrt{\epsilon/2}$,
the total energy per unit length ${\cal
E}/L=2\mathrm{\pi}\sqrt{2\epsilon}$, 
as easily follows from~(\ref{energy_expand}).
Therefore, the axial force $F/(2\mathrm{\pi})=\sqrt{2\epsilon}$
saturates to a constant value independent of tether length once
the tether has formed.  Since the axial force is independent of
$z$, the junction connecting the tether to the ring is a catenoid
with $b=\sqrt{2\epsilon}$.  Note that the minimum attainable
radius of the limiting catenoid is twice the radius of the tether.
Thus, the catenoid cannot smoothly join onto the cylindrical
tether and there must be a transition region (see Fig.~\ref{overlay_fig}).  
Before analyzing this
transition region in more detail, we consider the role of
pressure.

\subsubsection{The effects of pressure are subleading}
Evans and Yeung argued that the pressure jump $\Delta p$, present
in the case of a closed vesicle under tension, plays little direct
role in determining the tether radius~\cite{EY}.  Since a sphere
of radius $R_0$ has constant mean curvature $1/R_0$ and Gaussian
curvature $1/R_0^2$, the Euler-Lagrange equation~(\ref{eleqn}) in
the spherical region of the vesicle reduces to the Young-Laplace
law, $2H=\Delta p$ (even in the presence of bending resistance).
Measuring lengths in units of $R_0$ (for this paragraph only),
the Euler-Lagrange equation in the region of the tether
becomes
\begin{equation}
{\epsilon\over 2a^3}-{1\over a}+2=0, \label{pressure}
\end{equation}
since the pressure jump is everywhere uniform. For small
$\epsilon$, the three solutions are
\begin{eqnarray}
a&=&\pm\sqrt{\epsilon/2}+\epsilon/2+{\cal O}(\epsilon^{3/2})\\
a&=&1/2-\epsilon+{\cal O}(\epsilon^2). \label{rpressure}
\end{eqnarray}
The solution $a=-\sqrt{\epsilon/2}+{\cal O}(\epsilon)$ is
unphysical. The solution $a=\sqrt{\epsilon/2}+{\cal O}(\epsilon)$
corresponds to the tether in the case $\Delta p=0$. Thus, to
leading order, the tether radius is unchanged and the effect of
pressure appears at order $\epsilon$. Finally, the solution with a
radius near $1/2$ corresponds to a balance of pressure and
tension, and is not relevant for tethers.

\begin{figure}
\includegraphics[height=2.8in]{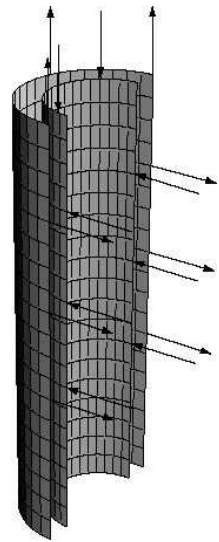}
\caption{Axial force due to the greater circumference of the outer
leaf.
} \label{tension_fig}
\end{figure}

Returning to our model problem with $\Delta p=0$, an apparent
paradox arises. Since an axial force is required to pull the
tether out of the membrane disc, there must be a tension in the
membrane. This tension is isotropic, since the membrane is fluid.
But consider the cylindrical portion of the tether between two
fixed values of $z$.  If this cylinder is cut in half along the
long axis (Fig.~\ref{tension_fig}), then each apparently
experiences a resultant force due to the tension. What force balances this
tension force if $\Delta p=0$?

The paradox is most readily resolved by comparing the
Euler-Lagrange equations of the variational approach with the
equations which follow from force and moment balance on a membrane
element, given the lipid bilayer membrane constitutive
relation~\cite{EY}.  
The appendix recapitulates the comparison between the two
approaches to the equilibrium shape equations. There it is shown
that the coefficient $\mu$ of the area term in the variational
energy is the tension only for minimal surfaces:
$\mu=\Sigma+\epsilon H^2$, where $\Sigma$ is the tension.  Since
the pressure jump $\Delta p=0$, the cylindrical region of the
membrane is in a state of pure bending, $\Sigma=0$.  But due to
the differential stretching inherent in bending, the outer sheet
is stretched and the inner sheet is compressed
(Fig.~\ref{tension_fig}). Since the sheets are fluid, the tension
or compression in each sheet is isotropic. The compressive and
tensile forces along the lines of longitude of the cylinder cancel
(consistent with $\Delta p=0$).  However, since the outer sheet is
longer than the inner sheet along a line of latitude, there is a
net axial tension.  The axial force at zero pressure jump is a
manifestation of the liquid properties of lipid bilayer membranes.

\subsubsection{The tether always necks twice}

We have seen that the shape of a membrane subject to a point force
and under high tension is best described as a boundary-layer
problem, with tension dominating in the outer region and bending
dominating in the inner region.  This situation is reminiscent of
the coating problem studied by Landau and Levich~\cite{LL}, and 
of Bretherton's bubble problem~\cite{B}, in which 
the shapes of different regions of
an interface are determined by different balances.  The analogy
goes further: Bretherton showed that the trailing edge of a large
air bubble, rising in a capillary tube filled with viscous liquid,
has a slight ripple~\cite{B}.  It will now be shown that there are slight
ripples in the shape of a lipid membrane at \textit{both} ends of
the cylindrical tether region.  These ripples have been noticed in
the numerical work of ref.~\cite{HBSZ}.

Since bending and tension are equally important in the
junction region, we must rescale the variables to balance these
two effects.  If $s_1$ is the arclength corresponding to a point
in the transition region, it is enough to assume   
that the radius $r(s_1)$ is small and close to
$\sqrt{\epsilon/2}$, without any further specification of $s_1$.  
It is therefore natural to rescale the
radius as in section V., $r=\rho\sqrt{\epsilon}$.  The further
rescalings $\sigma=(s-s_1)/\sqrt{\epsilon}$ and
$\zeta=(z-z(s_1))/\sqrt{\epsilon}$ lead to a balance of the
bending and tension terms:
\begin{equation}
\bar \nabla^2\bar H+2\bar H^3-2\bar H\bar K-\bar H=0,
\label{rescaled_eleqn}
\end{equation}
where
\begin{eqnarray}
\bar H&=&{1\over2}\bigg[{\rho_{\sigma\sigma}\over\zeta_\sigma}
-{\zeta_\sigma\over\rho}\bigg]\label{rescaled_meancurv}\\
\bar K&=&-{\rho_{\sigma\sigma}\over \rho}\label{rescaled_gausscurv}\\
\bar \nabla^2&=&{1\over \rho}{\mathrm{d}\over\mathrm{d}\sigma}
\rho{\mathrm{d}\over\mathrm{d}\sigma},\label{rescaled_laplacian}\\
\end{eqnarray}
and $\rho_\sigma^2+\zeta_\sigma^2=1$. Therefore, the transition
region is governed by the full nonlinear Euler-Lagrange equation,
and there are no further simplifications arising from the
smallness of $\epsilon$. However, one can use perturbation theory
to study the shape of the transition region near the tether.  Let
$\rho=1/\sqrt{2}+\eta$, with $\eta\ll1$. To leading order in
$\eta$, Eq.~(\ref{rescaled_eleqn}) becomes
\begin{equation}
\eta_{\sigma\sigma\sigma\sigma}+4\eta=0. \label{ripple_eqn}
\end{equation}
Note that Eq.~(\ref{ripple_eqn}) also follows immediately from
Eq.~(\ref{energy_expand}) with appropriate rescalings. There are
four independent solutions to Eq.~(\ref{ripple_eqn}), each of the
form $\eta_\alpha=C_\alpha \exp(ip\sigma)$, where $p=\pm(1\pm i)$
and $\alpha=1,...,4$. The shape near either end of the tether region
is an exponentially-damped sinusoid with wavelength
$2\mathrm{\pi}/{\sqrt\epsilon}$ and decay length
$\sqrt{\epsilon}$.

\begin{figure}
\includegraphics[height=3.in]{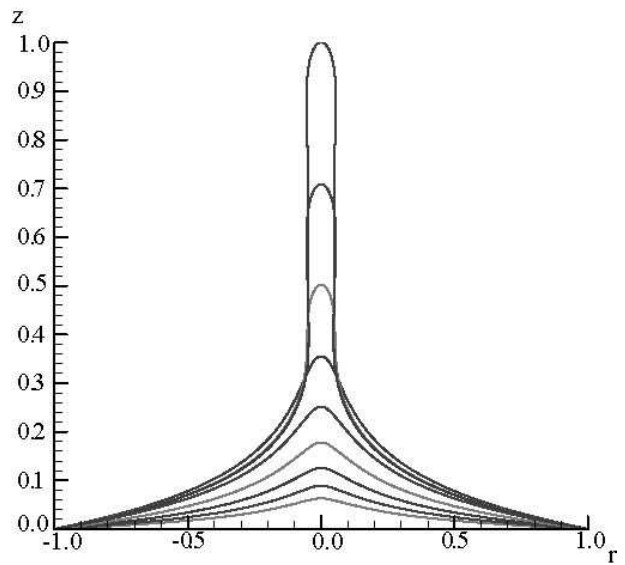}
\caption{Membrane shape for various ring separations;
$\epsilon=0.005$ and $r_0=0.005$.} \label{evolution_fig}
\end{figure}
\subsection{Tether:  numerical solution}

\begin{figure}
\includegraphics[height=3.in]{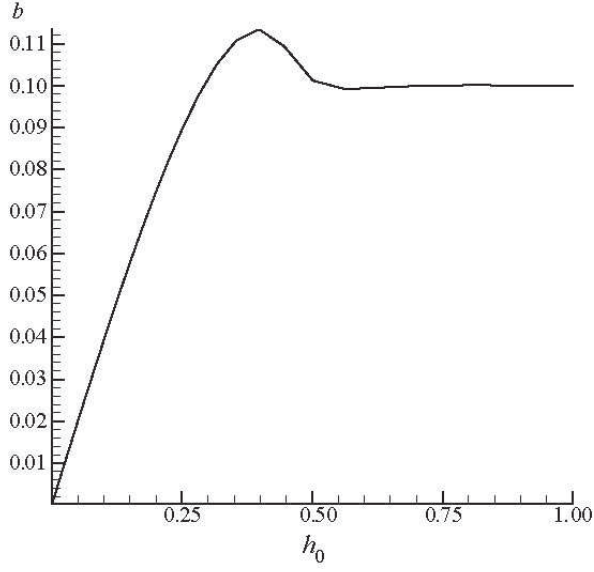}
\caption{Force \textit{vs.} displacement; $\epsilon=0.005$,
$r_0=0.005$.} \label{force_fig}
\end{figure}

\begin{figure}
 \includegraphics[height=3.in]{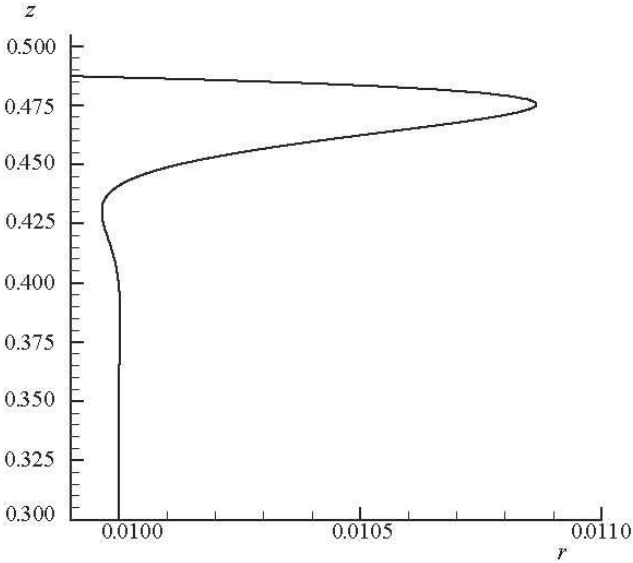}
 \caption{Ripple profile, $\epsilon=0.0002$.}\label{ripple_fig}
\end{figure}

The last section shows that a description of the
membrane shape in the junction region requires the solution of a
nonlinear differential equation with no small parameters, despite
the smallness of $\epsilon$.   Rather than solve this equation
numerically and match the solution onto the tether and catenoidal
regions, we simply solve for the complete shape numerically.  
Standard relaxation techniques~\cite{numerical_recipes} are used to
solve for the shape as a function of ring displacement $h_0$,
with a small ring of radius $r_0=0.001$ mimicking the point force.
Figure~\ref{evolution_fig} displays the membrane shape for various
$h_0$.  For small $h_0$, the shape is well approximated by the
linearized catenoid with an elastic boundary layer at small radius
(see section V.). As $h_0$ increases, the amplitude of the
catenoid increases until the limiting catenoid with
$b=\sqrt{\epsilon}$ is reached. For larger separations, a tether
forms. The axial force as a function of displacement is shown in
Fig.~\ref{force_fig}. Note that the force increases to a maximum
and then decreases slightly before saturating to
$\sqrt{2\epsilon}$.  This behavior is reflected in
Fig.~\ref{evolution_fig}, where the limiting catenoid lies inside
the catenoids with slightly lower values of $h_0$, since these
catenoids have slightly larger values of the minimum neck radius
$b=F/(2\mathrm{\pi})$.  Figure~\ref{ripple_fig} shows the ripple
in the junction region. The radial scale has been magnified for
clarity.

\section{Conclusion}

We have seen that tethers in our model problem are a type of
boundary-layer phenomenon.  In the cylindrical tether region,
bending dominates, whereas tension dominates at larger radii.  
These insights carry over to the more complicated problem
of tether formation in closed lipid bilayer membrane vesicles,
where the quantitative details of the force \textit{vs.} extension
will be different, since tension depends on extension. An important
generalization of the problem considered here would be to study
membranes with varying degrees of in-plane order, ranging from
liquid-crystalline to solid-like, since the liquid nature of fluid
membranes is crucial for tether formation.








\begin{figure}
\includegraphics[height=3.in]{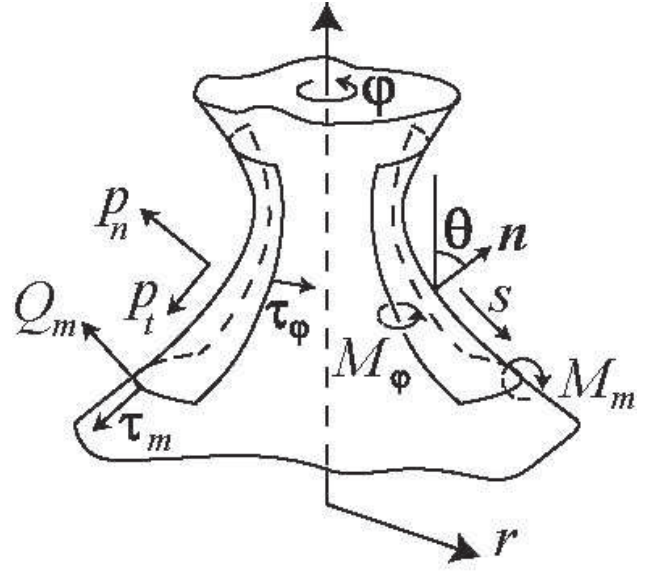}
\caption{Forces and moments acting on an element of an
axisymmetric membrane.} \label{element}
\end{figure}

\begin{acknowledgments}
We thank Vikram Deshpande for useful discussions and Deborah Fygenson 
for Fig.~\ref{deborahpicture}, and acknowledge
support from the Brown MRSEC on Micro- and Nanomechanics of
Materials (TRP) and NSF Grant DMR9812526 (REG and GH).
\end{acknowledgments}

\appendix
\section{Plate theory vs. variational principle}

This appendix reviews the force and moment balance relations
for axisymmetric shells~\cite{EY,love}, and the
constitutive relations for fluid membranes~\cite{EY}.  The
approach is equivalent to the variational approach taken in the
text, and elucidates the apparent paradox discussed in section
VB2. Figure~\ref{element} shows the forces and moments acting on a
small element of fluid membrane.  Only the forces and moments that
enter the shape equations are shown.  $\tau_m$ is the force per
unit length parallel to the meridian acting on an element edge
along the azimuthal direction.  $\tau_\varphi$ is the force per
unit length in the azimuthal direction acting on an element edge
along a meridian.  The shearing force $Q_m$ acts along the surface
normal on an element edge along the azimuthal direction.  The
external stresses $p_n$ and $p_t$ are forces per area acting on
the element in the normal and meridional directions, respectively.
The curvature along the meridian is
$c_m=\mathrm{d}\theta/\mathrm{d}s$, and the curvature in the
azimuthal direction is $c_\varphi=\sin\theta/r$.

The balance of forces and moments is just the same as in shells.
Normal stress balance requires
\begin{equation}
p_n=\tau_\varphi c_\varphi+\tau_m c_m-{1\over
r}{\mathrm{d}\over\mathrm{d}s}(rQ_m).\label{normal_stress}
\end{equation}
Tangential stresses balance when
\begin{equation}
-p_t={1\over
r}{\mathrm{d}\over\mathrm{d}s}(r\tau_m)-{\tau_\varphi\over
r}{\mathrm{d}r\over\mathrm{d}s}+c_m Q_m,\label{tangential_stress}
\end{equation}
where  $\mathrm{d}r/\mathrm{d}s=\cos\theta$. Moment
balance about the $\varphi$-axis relates the shearing force $Q_m$
to the moments per unit length $M_m$ and $M_\varphi$:
\begin{equation}
Q_m={1\over r}{\mathrm{d}\over\mathrm{d}s}(r M_m)-{M_\varphi\over
r}{\mathrm{d}r\over\mathrm{d}s}.\label{shearing_force}
\end{equation}

It is useful to have an expression for the total axial force
acting on a circle of latitude.  Consider the resultant of the
normal and axial stresses along the axial direction:
\begin{equation}
r(p_n\cos\theta-p_t\sin\theta)={\mathrm{d}\over\mathrm{d}s}
\bigg(r\tau_m\sin\theta-rQ_m\cos\theta\bigg),
\label{axial_stresses}
\end{equation} where we have used
Eqs.~(\ref{normal_stress},\ref{tangential_stress}). In our model
problem, the external stresses vanish, $p_n=p_t=0$. Thus,
$r\tau_m\sin\theta-rQ_m\cos\theta$ is a constant, 
the total axial force by $2\mathrm{\pi}$
(compare with the axial force on a soap
film, Eq.~(\ref{Noether})):
\begin{equation}
{F\over2\mathrm{\pi}}=r\tau_m\sin\theta-rQ_m\cos\theta.
\label{force}
\end{equation}

Returning to the derivation of the shape equations, consider now
the tangential forces per unit length $\tau_m$ and $\tau_\varphi$.
If $x$ denotes the coordinate across the thickness of the
membrane, then
\begin{eqnarray}
\tau_m&=&\bar\tau+c_\varphi M_m, \\
\tau_\varphi&=&\bar\tau+c_mM_\varphi, \label{stretch_const_reln}
\end{eqnarray}
where $\bar\tau_m=\int\tau_m\,\mathrm{d}x$,
$\bar\tau_\varphi=\int\tau_\varphi\,\mathrm{d}x$, $M_m=\int
x\tau_m\,\mathrm{d}x$, and $M_\varphi=\int
x\tau_\varphi\,\mathrm{d}x$. Part of the ``tension" in the membrane
comes from the bending moments.

The constitutive relation for the fluid membrane completes the
specification of the shape equations.  Since the fluid nature
implies isotropy, $\bar\tau_m=\bar\tau_\varphi$. Define the
common value of tension as $\tau=\bar\tau_m=\bar\tau_\varphi$.
Likewise,
\begin{equation}
M_m=M_\varphi=\kappa\bar c, \label{moment_const_reln}
\end{equation}
where $\bar c=c_m+c_\varphi$.
Thus, the shearing force is known once the curvature of the
membrane is known:
\begin{equation}
Q_m=\kappa{\mathrm{d}\bar c\over\mathrm{d}s}. \label{shear}
\end{equation}
Tangential force balance, Eq.~(\ref{tangential_stress}), becomes
\begin{equation}
-p_t={\mathrm{d}\over\mathrm{d}s}(\bar\tau+{1\over2}\kappa\bar
c^2). \label{tangential_forceII}
\end{equation}
In the absence of flow, the external tangential stresses vanish,
and $\tau\equiv\bar\tau+\kappa\bar c^2/2$ is constant. However,
$\bar\tau$ and $\bar c$ need not separately be constant.

Inserting the constitutive relations into
Eq.~(\ref{normal_stress}), the normal stress balance becomes
\begin{equation}
p_n=\tau\bar c-{1\over2}\kappa\bar
c(c_m-c_\varphi)^2-\kappa{1\over
r}{\mathrm{d}\over\mathrm{d}s}\bigg(r{\mathrm{d}\bar
c\over\mathrm{d}s}\bigg).\label{normal_stressII}
\end{equation}
But since $\bar c=2H$ and $K=c_mc_\varphi$,
Eq.~(\ref{normal_stressII}) reduces to the Euler-Lagrange
equation~(\ref{eleqn}) with $\mu=\tau$.

\end{document}